\documentclass[%
 reprint,
 preprintnumbers,
superscriptaddress,
 amsmath,amssymb,
 aps,
prb,
]{revtex4-1}

\usepackage{graphicx}
\usepackage{dcolumn}
\usepackage{bm}
\usepackage[hidelinks]{hyperref}
\usepackage[dvipsnames]{xcolor}
\usepackage{colortbl}
\usepackage{soul}
\usepackage{float}
\hypersetup{
    colorlinks,
    linkcolor={red!50!black},
    citecolor={blue!50!black},
    urlcolor={blue!80!black}
}

\begin{document}

\preprint{LA-UR-24-31680}

\title{Dynamical structure factors of Warm Dense Matter from Time-Dependent Orbital-Free and Mixed-Stochastic-Deterministic Density Functional Theory }

\author{Alexander J. White}
\email{alwhite@lanl.gov}
\affiliation{Theoretical Division, Los Alamos National Laboratory, Los Alamos, NM 87545, USA}

\noaffiliation

\date{\today}

\begin{abstract}
We present the first calculations of the inelastic part of the dynamical structure factor (DSF) for warm dense matter (WDM) using Time-Dependent Orbital-Free Density Functional Theory (TD-OF-DFT) and Mixed-Stochastic-Deterministic (mixed) Kohn Sham TD-DFT (KS TD-DFT). WDM is an intermediate phase of matter found in planetary cores and laser-driven experiments, where the accurate calculation of the DSF is critical for interpreting X-ray Thomson scattering (XRTS) measurements. Traditional TD-DFT methods, while highly accurate, are computationally expensive, motivating the exploration of TD-OF-DFT and mixed TD-KS-DFT as more efficient alternatives. We applied these methods to experimentally measured WDM systems, including solid-density aluminum and beryllium, compressed beryllium, and carbon-hydrogen mixtures. Our results show that TD-OF-DFT requires a dynamical kinetic energy potential in order to qualitatively capture the plasmon response. Additionally, it struggles with capturing bound electron contributions and accurately modeling plasmon dynamics without the inclusion of a dynamic kinetic energy potential. In contrast, mixed TD-KS-DFT offers greater accuracy in distinguishing bound and free electron effects, aligning well with experimental data, though at a higher computational cost. This study highlights the trade-offs between computational efficiency and accuracy, demonstrating that TD-OF-DFT remains a valuable tool for rapid scans of parameter space, while mixed TD-KS-DFT should be preferred for high-fidelity simulations. Our findings provide insight into the future development of DFT methods for WDM and suggest potential improvements for TD-OF-DFT.

\end{abstract}

\maketitle

\section{Introduction}\
Warm dense matter (WDM) is a complex, intermediate, phase between plasma, liquids, and solid.\cite{Bonitz2020, Dornheim2023-2,Falk2018} It occurs naturally in the cores of planets, in dwarf stars, and exoplanets. Artificially it is generated in both inertial confinement and high-powered laser experiments. \cite{Lorenzen2014,Prakapenka2021,Ehrenreich2020,Shawareb2024,Kraus22, Lütgert22, Hernandez2023, Roy2024} The temperatures can range from a few to hundreds of electron volts (tens of thousands to millions of Kelvin), at densities ranging from near ambient solid to multiple times compressed. The experimental generation and measurement of WDM has seen tremendous growth in the last two decades. \cite{Falk2018} One of the most important experiments for WDM is X-ray Thomson scattering (XRTS).\cite{Kraus_2016,Gawne2024} XRTS involves the measurement of scattered coherent hard xrays and directly probes the dynamic structure factor (DSF). Based on detailed balance, the DSF is related to the electron density-density response function.\cite{Gregori03,Dornheim2024} Combined with theoretical models of the response function, XRTS provides an approach to measure the electron density or ionization and/or the temperature of the WDM system.\cite{Nardi98,Lihua2012,Kraus2016,Doppner2023} While temperature can in principle be extracted directly, determination of ionization / electron density requires a reliable model for the DSF.\cite{Dornheim2023}

The total dynamic structure factor can be decomposed into elastic and inelastic parts. The elastic contribution yields the Rayleigh peak and is proportional to the ion-ion DSF, \textit{i.e.}, it is related to the dynamics of the ions and the equilibrium electron density. \cite{Gregori09,Vorberger2015,Dornheim2024-2} The inelastic contribution includes the dynamic response of the electrons, and generally contributes at higher frequencies.\cite{Diaw2017,Baczewski2016} The traditional approach to analyze XRTS is to utilize the Chihara decomposition which further separates the inelastic part into bound and free electron contributions. \cite{Chihara_2000} 

It is challenging to model the dynamics of warm dense matter. Coupling of the electrons, correlation and exchange, and Fermi degeneracy all play significant roles. Moreover, thermal excitations also have to be taken into account. Analytical kinetic theories typically become inaccurate in this regime. \cite{GRABOWSKI2020,Stanek2024} Mermin's finite temperature extension to Kohn-Sham density functional theory (Mermin-DFT) is a critical tool for modeling WDM as it can treat all these effects nonperturbatively.\cite{Mermin1965} The approximation in Mermin-DFT lies in the exchange-correlation free energy. Assuming the electrons thermalize on a shorter time-scale than the ionic motion, Mermin-DFT can be combined with molecular dynamics (MD) to simulate the ion-ion dynamic structure factor from first principles.\cite{Witte2017,Schorner2022,Schorner2022-2} Orbital-free density functional theory (OF-DFT) has been a complementary approach to Mermin-DFT for the calculation of equation of state, \cite{GAFFNEY2018,Hu16,Hu17,Ding17,Hu2018} ionic transport,\cite{Ticknor2016, White2017,White2019,GRABOWSKI2020,Clerouin2020,Ticknor2022} and structure factors.\cite{Mi2023, T-White2013, Xu2024} It is based on an approximate noninteracting kinetic free-energy density functional. The direct dependence of the kinetic energy from the electron density, rather than the auxiliary Kohn-Sham orbitals, makes OF-DFT much more efficient (by several orders of magnitude). While OF-DFT is known to miss ``shell-effects", it is still a reasonably accurate approach for the simulation of WDM, and considerable efforts to develop more accurate, often nonlocal, kinetic energy density functionals continue. \cite{Wang1992,Sjostrom2014,Sun23,Lüder_2024,Sun24, Vargas2024,Dieterich2017}

To model the electronic time-scale dynamics, which yields the inelastic part of the DSF, requires a different approach. Following the Chihara decomposition, \cite{Chihara_2000} the free electron contribution can be modeled via the response function of a free homogeneous electron gas using the random phase approximation, local field corrections and/or collision frequencies.\cite{Plagemann_2012,Schorner23} But the separation between bound and free electrons is not always clear, and the treatment of bound-free transitions still requires explicit calculation.\cite{Plagemann_2012,Johnson2012,Mattern2013, Hou2015} Thus, recently the separation the partitioning of the inelastic DSF has been avoided by calculating the electronic response directly from time-dependent Mermin-DFT (TD-DFT).\cite{Baczewski2016} The free-free and bound-free contributions can then be inferred from multiple calculations using pseudo-potentials which fix different numbers of ``frozen-core" electrons.\cite{Baczewski2016} However, TD-DFT is often a computationally prohibitive approach, particularly at higher temperatures and/or lower densities.

While its roots trace back over 40 years, the extension of the OF-DFT to treat electron dynamics, Time-dependent OF-DFT (TD-OF-DFT), has only recently been caught-on as a much more efficient alternative to TD-DFT. \cite{Ball73, Domps1998,Palade_2015} It has been used successfully to calculate the stopping power of warm dense matter in good agreement with experimental and TD-KS-DFT results.\cite{White2018,Ding2018,Malko22} It has also been used to calculate the absorption spectrum of plasmonic nanoparticles and ion clusters.\cite{Xiang2020, Covington2021, Jiang2021, Jiang2022, Sala2022} The approach is a reformulation of a hydrodynamic treatment of the electron density, under the influence of a Fermi and exchange-correlation pressure.\cite{Moldabekov2018, White2018, Graziani2022} Alternatively it can be viewed as a Bosonic treatment of electrons, \textit{i.e.}, with a single condensate wavefunction, under the influence of an (approximated) effective Fermi / Pauli potential.\cite{White2018, Jiang2021} Additional dynamical kinetic energy potentials have been developed, by relating the approximate TD-OF response of a free electron gas to the exact response.\cite{Neuhauser2011,White2018,Jiang2021,Jiang2022} Similar in practice, quantum hydrodynamic models including an electron viscosity have also been proposed. \cite{Rusek2000,Michta2015, Diaw2017}

Here, we investigate the application of TD-OF-DFT to the calculation of the inelastic part of the DSF by comparing TD-OF-DFT results to Mermin-Kohn-Sham based TD-KS-DFT. We utilize the mixed-stochastic-deterministic (MSD) approach to TD-DFT, which significantly increases the efficiency of TD-DFT simulations of warm dense matter, while retaining the accuracy of traditional Kohn-Sham TD-DFT. \cite{White2020,White2022} However, we note that TD-OF-DFT is still much more efficient than mixed TD-KS-DFT. We consider four experimentally measured systems which span the warm dense matter regime, solid-density Aluminum (Al)  at $k_B T=6$ eV,\cite{Sperling2015} solid-density Beryllium (Be) at $k_B T=53$ eV,\cite{Glenzer2003} and  6.05 $\text{g/cm}^3$ Carbon-Hydrogen mixture (CH) at $k_B T=10$ eV;\cite{Fletcher2014} and to verify our approach we also consider the 5.5 $\text{g/cm}^3$ $k_B T=13$ eV Be case which has previously been calculated using real-time TD-DFT and reported. \cite{Lee2009,Baczewski2016}

\section{Methods}

\subsection{ Linear Response Dynamic Structure Factor from the Time-Domain}
The fluctuation dissipation theorem relates the inelastic part of the Dynamic structure factor $S_{\bf{q}}(\omega)$ is related to the imaginary part of the density-density response function, $\chi_{{\bf{q}},{\bf{q}}}(\omega)$.\cite{Sturm1993} Note that for the remainder of the manuscript we utilize atomic units, thus $\{k_B, \hbar, m_e, e, 4 \pi \epsilon_0 \} \equiv 1$. 
\begin{align}
\label{eq:Sqw}
   S_{\bf{q}}  (\omega) = -\frac{1}{\pi } \frac{\Im\{\chi_{\bf{q},\bf{q}}(\omega)\}}{1-e^{-\omega/{T}}}  
\end{align}

From linear response theory the general density-density response function is given by:
\begin{align}
\label{eq:Xqw}
   \delta {\tilde \rho}_{\bf{q}}(\omega) &= \chi_{{\bf{q}},{\bf{q}}'} (\omega) \delta {\tilde V}_{{\bf{q}}'}( \omega) \nonumber 
   \\ \to 
    \chi_{{\bf{q}},{\bf{q}}'} (\omega)  &= \delta {\tilde \rho}_{\bf{q}}(\omega) / \delta {\tilde V}_{{\bf{q}}'}( \omega) .
\end{align}

$\chi_{{\bf{q}},{\bf{q}}}(\omega)$ can be calculated using real-time dynamics by Fourier transforming the intermediate density response ${ \delta \rho}_{{\bf{q}}} (t) $ which is responding to a perturbation which is nearly independent of frequency and a delta-function of the targeted wave vector (${\bf{q}}$), \emph{i.e.},
\begin{align}
\label{eq:Xqw}
    \nonumber
   \delta {V}({\bf{\hat r}}, t) =  v_0 g(t) e^{i {\bf{q}} \cdot {\bf{ \hat r}}}, \\ 
   \delta { \rho}_{{\bf{q}}} (t) =  \text{Tr}\{ e^{i {\bf{q}} \cdot {\bf{ \hat r}}} \,   \big( \hat{\rho}(t) - \hat{\rho}(0) \big) \}, \text{and} \\
   {\tilde V}_{{\bf{q}}}(\omega) =  v_0 g(\omega) \delta_q 
\end{align}

For isotropic systems we can exploit the equivalence of the $\pm {\bf{q}}$ responses and consider only real perturbation and observable, \textit{i.e.}, we replace $e^{i {\bf{q}} \cdot {\bf{r}}}$ with $ 2\cos ({\bf{q}} \cdot {\bf{r}})$.\cite{Groth2019} $g(t)$ describes the time-dependence of the perturbation. Its Fourier transform must be non-zero for all non-zero frequencies in $\chi_{{\bf{q}},{\bf{q}}}(\omega)$. It is chosen to be a sharp, relative to the dynamics of the response, normalized Gaussian envelope. $v_0$ is a constant which is chosen to be large enough to minimize numerical noise, and small enough to avoid non-linear effects.\cite{Baczewski2016} $\hat \rho(t)$ is the time-dependent density matrix. 

Thus-far we have described general linear-response theory. The electronic structure method used to perform the calculation determines the time-dependent density matrix $\hat \rho$(t). We will now describe the electronic structure methods compared in this article. 

\subsection{Finite-Temperature Equilibrium DFT}

\subsubsection{Mermin-Kohn-Sham DFT}

In the Mermin-Kohn-Sham approach the $t=0$, equilibrium, single-particle $\hat \rho$ is given as the Fermi-Dirac (FD) weighted Kohn-Sham orbital outer-product (in Dirac notation):
\begin{align}
\label{eq:DMKS}
    {\hat \rho_{KS}} = \sum_a f(\varepsilon_a, \mu, T) \, \vert \phi_a \rangle \langle \phi_a \vert.
\end{align}
where the orbitals are eigenvectors of the Kohn-Sham Hamiltonian:
\begin{align}
\label{eq:KS}
    {\hat H_{KS}} = -\frac{\nabla^2}{2} + V_{xc}({\bf{r}},[\rho]) &+ V_{ext}({\bf{r}})  + \int d^3{\bf{r'}} \frac{{\hat \rho}({\bf{r'}},{\bf{r'}})}{\vert {\bf{r'}} - {\bf{r}}\vert} \nonumber 
    \\
    {\hat H_{KS}} \vert \phi_a \rangle   &= \varepsilon_a \vert \phi_a \rangle 
    \nonumber 
    \\
    f(\varepsilon, \mu, T) &= \frac{2}{1+e^{(\varepsilon-\mu)/{T}}}  
\end{align}
The factor of 2 accounts for our spin-unresolved treatment of the electrons. Eqs \ref{eq:DMKS} and \ref{eq:KS} are solved self-consistently. For $T=0$ the FD function is a step function which limits the rank of the density matrix, Eq. \ref{eq:DMKS}, to half the number of electrons. At non-zero temperatures the rank of ${\hat \rho_{KS}}$ is formally equal to the basis size, but in practice is safely truncated by neglecting orbitals with occupation, $f(\varepsilon, \mu, T)$, less than $\sim 10^{-5}$. As temperature becomes high the Fermi-Dirac function becomes flat and extends to high energies, $\sim 5 \times T$, such that the total number of states required to converge ${\hat \rho_{KS}}$ approximately scales as $N_\phi \approx N_{e}\times \{1+({T/T_{F}})^{3/2} \}$, where $T_F$ is the Fermi temperature, and $N_e$ is the number of electrons in the system.\cite{Blanchet2020} The computational cost of solving the eigenproblen in Eq. \ref{eq:KS} scales as $N_B N_\phi^2$ ($N_B$ is the computational basis size) leading to cubic scaling with both system size and temperature. This computational complexity led to the development of alternative schemes for determining ${\hat \rho_{KS}}$. 

\subsubsection{Stochastic DFT}
Stochastic DFT was developed as a linear-scaling (with respect to system size) alternative to the traditional approach just described.\cite{Gao2015,Cytter2018} It avoids solving the eigenproblem and instead works directly with the density matrix. The problem is kept tractable by working with a stochastic, rather than orbital-based, resolution of the identity matrix (ROI). 
\begin{align}
\label{eq:DMStoc}
     {\hat I} &= \sum_a \vert \phi_a \rangle \langle \phi_a \vert \to  \sum_\alpha^{N_S \to \infty} \vert \xi_\alpha \rangle \langle \xi_\alpha \vert
     \\
    {\hat \rho_{KS}}& = \sum_a f^{1/2}({\hat H_{KS}}, \mu, T) \, \vert \phi_a \rangle \langle \phi_a \vert \, f^{1/2}({\hat H_{KS}}, \mu, T)    \nonumber 
    \\
    \to  &\sum_\alpha^{\infty} f^{1/2}({\hat H_{KS}}, \mu, T) \, \vert \xi_\alpha \rangle \langle \xi_\alpha \vert \, f^{1/2}({\hat H_{KS}}, \mu, T) \nonumber 
    \\
    & \equiv \sum_\alpha^{\infty} \vert \Xi_\alpha \rangle \langle \Xi_\alpha \vert 
\end{align}
In real space the random vectors are formulated as :
\begin{align}
\label{eq:randvec}
      \langle {\bf{r}} \vert \xi_\alpha \rangle \equiv \frac{1}{\sqrt{N_S d^3{\bf{r}}}} e^{i2\pi\theta({\bf{r}})} \text{ or} \nonumber\\
      \langle {\bf{r}} \vert \xi_\alpha \rangle \equiv \frac{1}{\sqrt{N_S d^3{\bf{r}}}} \pm({\bf{r}}).
\end{align}
When complex vectors are utilized (non-$\Gamma$ point an/or TD-DFT) $\theta({\bf{r}})$ is a random number between 0 and 1, independently drawn for each position. For molecular dynamics at $\Gamma$, where real functions are prefered, $\pm({\bf{r}})$ is a randomly drawn $\pm 1$ at each position.  The stochastic convergence of the ROI error goes as $N_S^{-1/2}$. The computational bottleneck in stochastic DFT calculations is the ``filtering" of the stochastic vectors, \textit{i.e.}, calculating $\vert \Xi_\alpha \rangle = f^{1/2}({\hat H_{KS}}, \mu, T) \, \vert \xi_\alpha \rangle$. This involves the matrix function expansion of $f^{1/2}$, which scales as $E_{cut}/T$. \cite{Cytter2018} This scaling is advantageous for hot matter, but the prefactor is high. \cite{White2020,Sharma2023}
We note that the ROI approximation above is different from the utilization of the stochastic vectors, Eq. \ref{eq:randvec}, directly as a basis. The latter would involve the stochastic ROI applied to both the columns and vectors of $\hat H_{KS}$ and subsequent generation of a density matrix in that reduced subspace. This would likely not be very accurate for a reasonable number of vectors. 

\subsubsection{Mixed Stochastic-Deterministic DFT}
To improve the precision of the stochastic DFT approach, while keeping computational cost controlled we developed the mixed stochastic deterministic DFT approach.\cite{White2020,Sharma2023,White2022} In this approach the identity matrix is partitioned into a low-rank projector subspace and it's complement.
\begin{align}
\label{eq:IdenMixed}
     {\hat I} &= {\hat P} + \{ {\hat I} - {\hat P} \} \equiv {\hat P} + {\hat Q}
\end{align}
Since ${\hat P}$ is a projector, ${\hat P}={\hat P}^2$, $\hat Q$ is also a projector ${\hat Q}={\hat Q}^2$. Using a subset of Kohn-Sham orbitals to formulate ${\hat P}$ immediately leads to the mixed stochastic-deterministic Kohn-Sham density matrix
\begin{align}
\label{eq:DMMixed}
    {\hat \rho_{KS}}& = \sum_a^{N_\phi} f(\varepsilon_a, \mu, T) \, \vert \phi_a \rangle \langle \phi_a \vert \, +   
    \\
    &\sum_\alpha^{N_s \to \infty} f^{1/2}({\hat H_{KS}}, \mu, T) \, \vert {\tilde \xi}_\alpha \rangle \langle {\tilde \xi}_\alpha \vert \, f^{1/2}({\hat H_{KS}}, \mu, T) \nonumber 
    \\
    & \equiv \sum_a ^{N_\phi} f(\varepsilon_a, \mu, T) \, \vert \phi_a \rangle \langle \phi_a \vert \, + 
     \sum_\alpha^{N_s \to \infty} \vert {\tilde \Xi}_\alpha \rangle \langle {\tilde \Xi}_\alpha \vert  \nonumber,
     \\
     &\text{where} \, \vert {\tilde \xi}_\alpha \rangle \equiv \{ {\hat I} - \sum_a^{N_\phi} \vert \phi_a \rangle \langle \phi_a \vert \} \vert {\xi}_\alpha \rangle
\end{align}
The mixed approach recovers the traditional deterministic and stochastic approaches in the limits that $N_s \to 0$ or $N_\phi \to 0$ respectively. For a fixed, $N_\phi$, the stochastic precision converges at the same rate, but the prefactor decreases with increasing $N_\phi$.\cite{White2020,Sharma2023} Thus since the lowest energy eigenpairs are easiest to calculate, and have the largest occupations, it is an effective means to improve the precision without significantly increasing the cost. However, the computational cost is still comparable to zero-temperature Kohn-Sham DFT, and considerable when compared to more approximate methods.

\subsubsection{Orbital-Free DFT}
In orbital-free DFT, the density matrix is not usually considered. The energy of the system is given as a direct functional of the real-space density, $E_{OF}{[\rho]}$, alone. \cite{Xu2024} However, practically we can formulate OF-DFT in terms of a rank-one density matrix. 
\begin{align}
\label{eq:DMMixed}
    {\hat \rho_{OF}} &= N_e \, \vert \Psi  \rangle \langle \Psi \vert \\
    \text{where}  \, \langle {\bf{r}} \vert \Psi \rangle &= \sqrt{\rho({\bf{r}},{\bf{r}})} / \sqrt{N_e} \nonumber 
    \\
    \text{and}  \,  \frac{\partial E_{OF}}{\partial \rho({\bf{r}},{\bf{r'}}) }  \cdot \langle {\bf{r'}} \vert \Psi \rangle  \nonumber &= \mu_{OF} \langle {\bf{r}} \vert \Psi \rangle
\end{align}
This resembles the density matrix of a zero-temperature Bose-Einstein Condensate with $\vert \Psi  \rangle $ resembling a normalized Madelung wavefunction.\cite{White2018,Jiang2021,Jiang2022} The Orbital-Free energy and Hamiltonian must then include additional physical effects such as thermal excitation and Fermi statistics, which is lost in this dramatic reformulation. We will discuss the effect this has on electron dynamics in the following section. For the equilibrium case, the energy functional must include an extra kentropic contribution to the free energy.     
\begin{align}
\label{eq:HOF}
    {\hat H_{OF}} = &V_{xc}({\bf{r}},[\rho]) + V_{ext}({\bf{r}})  + \int d^3{\bf{r'}} \frac{{\hat \rho}({\bf{r'}},{\bf{r'}})}{\vert {\bf{r'}} - {\bf{r}}\vert}  \nonumber \\ 
     &-\lambda\frac{\nabla^2}{2} +  \gamma{\hat V}_{OF}([\rho], T)
\end{align}
${\hat V}_{OF}([\rho])$ is the, possibly non-local, orbital-free potential. The simplest orbital-free potential which introduces degeneracy and thermal effects comes from the Thomas-Fermi-Perrot kentropic functional, ${\hat E}_{TFP}([\rho])$. $\lambda$ and $\gamma$ are a real weights for the Von-Weizsäcker and local kinetic contributions, which allows for potential empirical tuning. In previous work it has been shown that taking $\gamma=1$ and $\lambda = 1$ yields the correct high and low wavevector limits of the static response function for a noninteracting homogeneous electron gas.\cite{Wang1992, Sjostrom2014, White2018, Moldabekov2018} We call this the Thomas - Fermi - Von Weizsäcker approach, ${\hat H_{TFW}}$.

\subsection{Finite-Temperature Time-Dependent DFT}
While we are calculating a linear response property, we choose to utilize real-time TD-DFT. This is convenient in the WDM regime, because of it is simple and it eliminates the ``sum-over-states" expansions, thus it scales only quadratically with system size for fully deterministic Kohn-Sham DFT and linearly for stochastic and orbital-free DFT. In the real-time approach to time-dependent DFT the density matrix has to evolve in time according to the Liouville–von Neumann equation.
\begin{align}
    i\frac{\partial}{\partial t} {\hat \rho} (t)  = [{\hat H} (t) , {\hat \rho}]
\end{align}
Since we have formulated the OF and KS DFT approaches in terms of Hermitian low-rank density matrices, we can propagate the vectors (orbitals, filtered stochastic vectors, or Madelung-like) via using their respective Hamiltonians starting from their equilibrium results. 
For each vector then we have a Schr{\"o}dinger-like equation:
\begin{align}
    i\frac{\partial}{\partial t} \vert {\psi (t) } \rangle = {\hat H} (t)  \vert {\psi (t) } \rangle, 
\end{align}
with $ \psi \to \{ \phi, \Xi, \text{ or } \Psi\}$ and ${\hat H} \to \{ {\hat H_{KS}} \text{ or } {\hat H_{TFW}}\}  $. 
We utilize the explicit short iterative Lanczos approximation to propagate the vectors. \cite{Park1986} 
\begin{align}
   \vert \psi(t+\delta t) \rangle = e^{i{\hat H}[{\hat \rho}(t), t] \delta t} \vert {\psi(t)} \rangle, 
\end{align}
The density matrices described in the previous section define our initial vectors and their weights. These weights do not change during time-dependent propagation in this approach, the vectors do. This means that no unoccupied vectors need to be considered. 

\subsubsection{Dynamic Kinetic Energy Potential for time-dependent orbital-free DFT}
As described the KS and OF-DFT fall within the \textit{adiabatic} TD-DFT approximation, meaning the respective Hamiltonians only depend on the time-local electron density. 

For Kohn-Sham TD-DFT the adiabatic approximation neglects \textit{dynamical} correlation and exchange. For orbital-free DFT additional noninteracting dynamical Fermi / thermal effects are also neglected. These latter effects emerge from the many-electron / high-rank form of the Kohn-Sham density matrix, but need to be explicitly added to the orbital-free Hamiltonian. Following the same logical approach as used to derived the static TFW potential, we previously proposed a simple time-local current-density dependent kentropic potential.\cite{White2018} When added to the TFW Hamiltonian (TFW+D) it recovers all high and low combinations of the $q$ and $\omega$ dependent \textit{dynamical} noninteracting free electron gas response, \textit{i.e.}, the Lindhard response, $\chi_L(q,\omega)$. The TFW response of a HEG at $T=0$ is given by:
\begin{align}
   \chi_{TFW} (q, \omega) = -\frac{k_F}{\pi^2}\big\{1+\frac{3q^2}{4k_F^2}-\frac{3w^2}{4k_F^2q^2}\big\}^{-1}
\end{align}
where $k_F$ is the Fermi momentum of the HEG. The TFW+D improves the model by adding the first dynamical, $\propto \omega$, term in the small $q$ expansion of the inverse response function.  
\begin{align}
   \chi_{TFW+D} (q, \omega) = -\frac{k_F}{\pi^2}\big\{1+\frac{3q^2}{4k_F^2}-\frac{3w^2}{4k_F^2q^2} + i\frac{\pi\omega}{2k_Fq}  \big\}^{-1}
\end{align} 
From the continuity equation, and assuming a ``local" Fermi momentum we generate our zero-temperature dynamical potential as: 
\begin{align}
   V_{Dyn} ({\bf{r}}, t, T=0) =   \frac{\pi^3}{4E_F(\bf{r})} \mathcal{F}^{-1}\big[\frac{i\bf{q}}{\vert \bf{q} \vert } \cdot {\tilde J}(\bf{q}, t) \big] (\bf{r}),
\end{align} 
where $E_F(\rho({\bf{r}},t))$ is the time and space local Fermi energy, and ${\tilde J}(\bf{q}, t)$ is the Fourier transform of the time-dependent electron current. Temperature is included through a local scaling factor which depends on the local degeneracy parameter. \cite{White2018}
\begin{align}
   V_{Dyn} ({\bf{r}}, t,T) &=    C_T ({\bf{r}}, t,T) \times V_{Dyn}({\bf{r}}, t,T=0) \nonumber \\
   C_T ({\bf{r}}, t,T) &= \big( [\theta({\bf{r}},t,T) \times 1.69271]^{3.6} +1 \big)^{\frac{1}{3.6}} \nonumber \\
   \theta({\bf{r}},t,T) =& \frac{T}{E_F(\rho({\bf{r}},t))}
\end{align} 
The main effect of this dynamical potential is to give a finite width to the HEG response pole and thus dissipate finite-$\bf q$ perturbations in real time. Higher order expansion terms in both $\omega$ and $\bf q$ have been utilized to improve TD-OF-DFT results in the frequency and time-domain. \cite{Jiang2021,Jiang2022} However, since the TFW+D approach recovers already the large $\omega$ and $\bf q$ limits and the peak position of the high $\omega$ dispersion curve of the HEG, we do not include any terms which alter these limits. Linear or higher $\bf q$ terms in the low $\omega$ expansion of $\chi_{Lind}^{-1}$ leads to worse agreement in the derivative of the imaginary response with respect to $\omega$ (the dampening term) in the high $\bf q$ limit. Moreover, the temperature dependence of each new term would need to be numerically determined, which is nontrivial. \cite{White2018} It should also be noted that the introduction of the dynamical potential results in a non-conservative force on the electron density and the resulting TD-OF-DFT dynamics do not conserve energy. This is not a significant issue, as we are applying linear response perturbations and treat the electrons as being in equilibrium, \textit{i.e.} the energy of the perturbation is negligibly small compared to the kinetic energy of the electrons which act as a thermal reservoir.

\subsection{Molecular dynamics generation of snapshots}

Our goal is to compare only the electronic response (the inelastic part of the DSF) calculated with TD-OF-DFT and TD-DFT. We therefore perform our TD-DFT calculations on the same set ionic positions (snapshots) for all cases. Those snapshots are generated via iso-kinetic molecular dynamics simulations,\cite{Minary:2003} performed with mixed-DFT.\cite{White2020,Sharma2023} We average over 5 snapshots for each calculation, with those snapshots chosen by taking our MD trajectory, eliminating a equilibrium period, then maximizing the time between each snapshot. The time between snapshots is more than sufficient for the positions to be uncorrelated.

\section{Results}

\begin{figure}[t]
    \centering
    \includegraphics[width=0.85\linewidth]{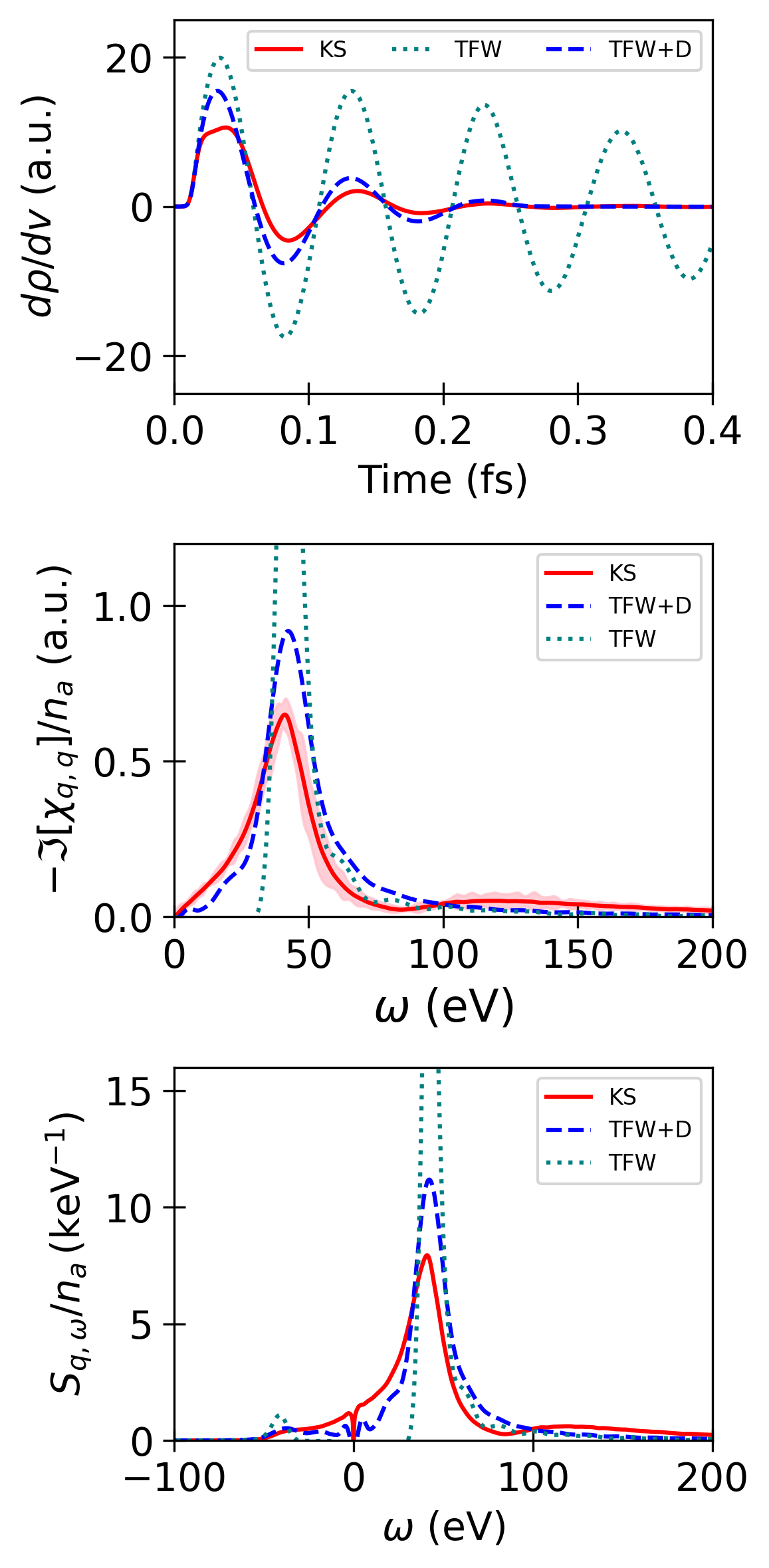}
  \caption{ Compressed Beryllium at density $\rho=5.5$ $\text{g/c
m}^3$ at $T=13$ eV. Results are averaged over five molecular dynamics snapshots with 64 atoms. Results are calculated using Kohn Sham (KS, red solid), Orbital Free at the Thomas Fermi von-Weizsäcker level (TFW, teal dotted), and Orbital-Free with TFW plus dynamic potential (TFW+D, blue dashed) (Top) Time-dependent density response at perturbing wavevector, $q=1.09\, {\mathring{\text{A}}}$, in atomic units. (Center) Negative of the imaginary part of the density response function per atom. Shaded band marks $\pm 10 \times$ the standard error, from both stochastic vector and MD sampling. (Bottom) The inelastic dynamic structure factor resulting from the response.}
    \label{fig:Be_1}
\end{figure}

We compare Kohn-Sham and Orbital-Free TD-DFT for calculation of the inelastic dynamic structure factor of experimentally measured warm dense matter systems.\cite{Sperling2015,Glenzer2003,Fletcher2014,Lee2009} We use both the TFW and the TFW+D. Our Kohn Sham calculations are done using mixed stochastic deterministic DFT. The Stochastic and Hybrid Representation of Electronic structure by Density functional theory (SHRED) code is used to perform the calculations. Further details and a tabulation of the simulation parameters can be found in the supplementary materials.\cite{Perdew96,Goedecker1996,Hamann2013,Setten2018,Monkhorst1976}

We consider 5.5 $\text{g/cm}^3$ Beryllium at 13 eV as our first case.\cite{Lee2009} We use this simulation to verify our approach, as it was already simulated using Kohn-Sham DFT by Backzewski \textit{et. al.} in the first calculations of the inelastic DSF of warm dense matter by TD-DFT . Our KS calculations for $q=1.09\, {\mathring{\text{A}}}$ show excellent agreement with their previous result. In the time-domain the density response of the KS and TFW+D appear qualitatively similar, while the TFW response is clearly underdamped. This is exactly what we would expect based on the role of the dynamical potential. The dampening rates for the KS and TFW+D appear to agree, though the KS result clearly has more than one characteristic frequency.     
In the frequency domain, the KS plasmon signal peaks at $\sim ~40$ eV with the 1s to continuum excitation occurring starting at $\sim 100$ eV. We show a magnified, $10\times$, standard error band around the KS result. This standard error results from both the stochastic vector and snapshot sampling. The TFW$+$D approach does not distinguish the character of the 1s and 2s electrons, merging the peaks into one slightly up-shifted, slightly broader, and larger magnitude plasmon peak. This is however, much improved compared to the TFW result. The width of the plasmon response is purely due to artificial broadening, $\Gamma = 5 $ eV required to prevent aliasing effects in the Fourier transform. Already this example calculation shows the qualitative differences of KS, TFW+D and TFW approaches. However, the temperature is relatively low and we would not necessarily expect high accuracy from OF methods. 

\begin{figure}[t]
    \centering
    \includegraphics[width=0.85\linewidth]{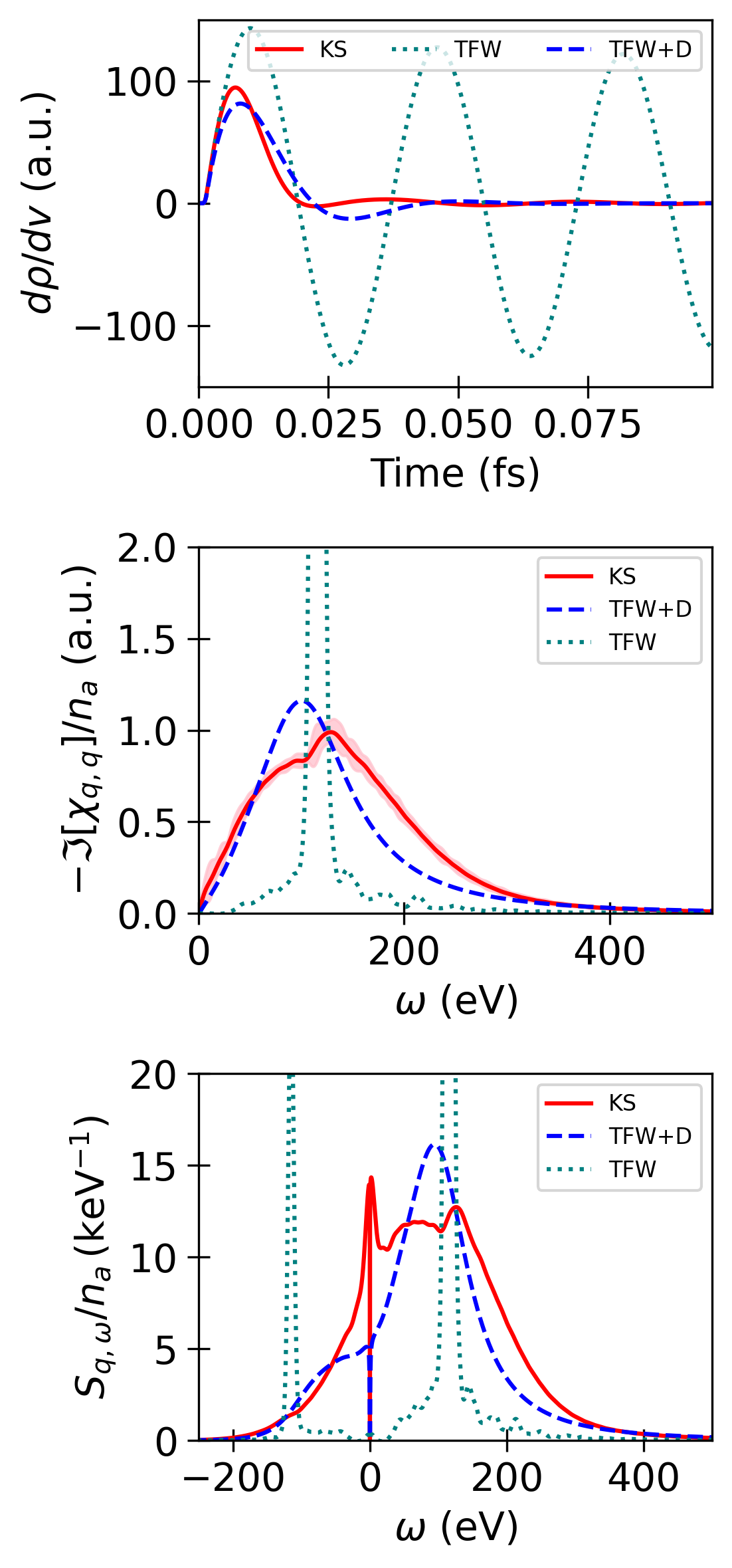}
  \caption{ Solid density Beryllium at $\rho=1.86$$ \text{g/cm}^3$ at $T=53$ eV. Same format as Fig.\ref{fig:Be_1}. Results are averaged over five molecular dynamics snapshots with 64 atoms.   }
    \label{fig:Be_2}
\end{figure}

\begin{figure}[t]
    \centering
    \includegraphics[width=0.85\linewidth]{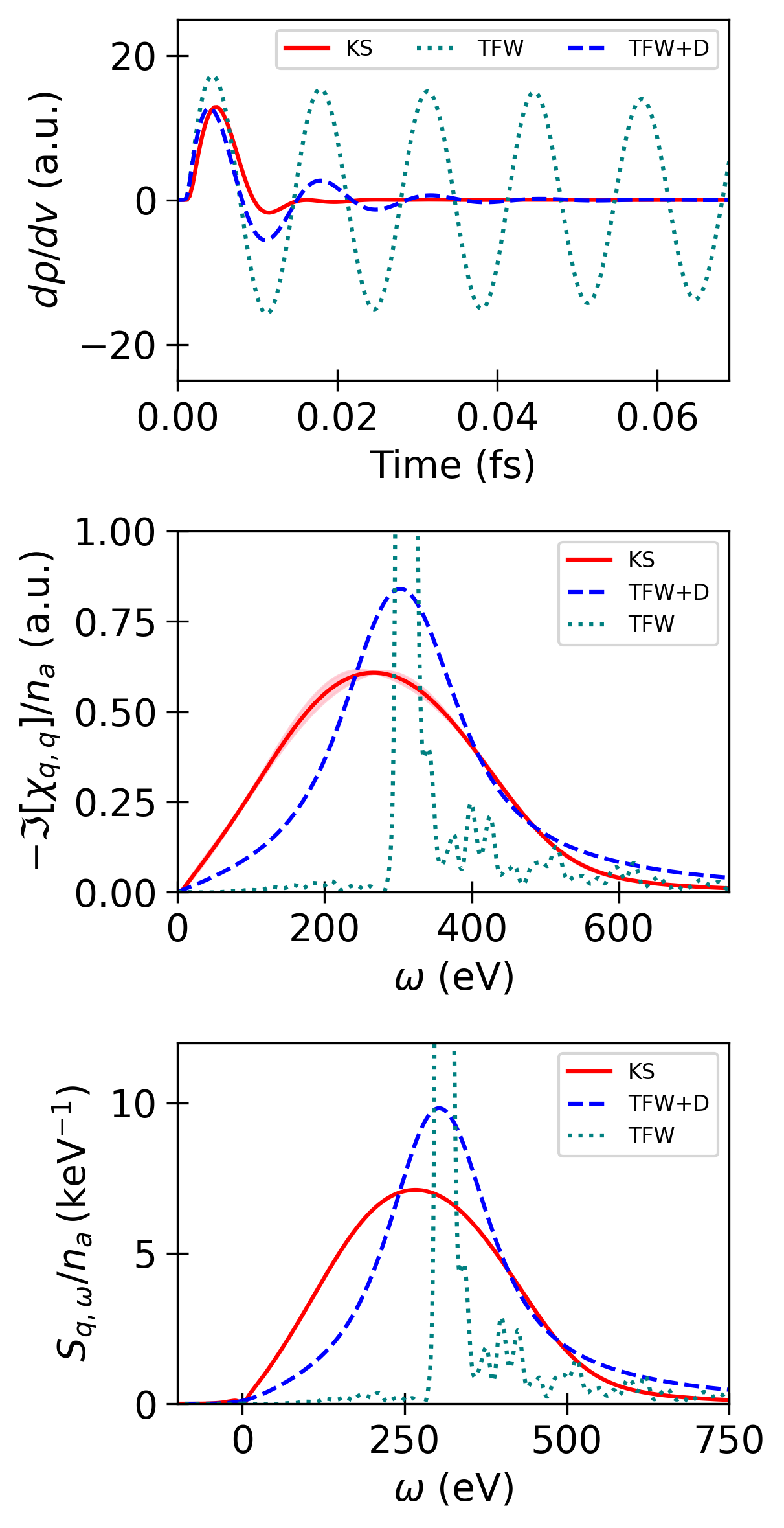}
  \caption{ 1:1 C:H mixture at $\rho=6.05$ $\text{g/cm}^3$ at $T=10$ eV. Same format as Fig.\ref{fig:Be_1}. Results are averaged over five molecular dynamics
snapshots with 32 C and 32 H atoms. }
    \label{fig:CH}
\end{figure}

The 2003 XRTS experiment of Glenzer \textit{et. al.} was the first of X-ray scattering, $q=4.27\, {\mathring{\text{A}}}$,  measurement of solid density plasma.\cite{Glenzer2003} Their analysis estimated a temperature of 53 eV. This is much higher than the previous case, and well within the range where we would expect for OF-DFT molecular dynamics to perform well. Fig. \ref{fig:Be_2} shows the results, similar format to Fig. \ref{fig:Be_1}, for this higher temperature system. We see qualitatively similar behavior to the previous higher density lower temperature case. T The TD-OF-DFT (TFW+D) still generates essentially a single plasmon response, while the KS result shows a peak due to the 1s orbitals. However, the KS result shows that the plasmon response is much broader, leading to overlap with the 1s to continuum transitions. The 1s orbitals are also partially ionized. Thus the agreement is only marginally better than the cooler and denser case. Without the dynamic potential the TD response essentially does not decay on a physical timescale. On a positive note the dynamic potential captures the order-of-magnitude decrease in the decay time for this response compared to the previous Be case (Fig.\ref{fig:Be_1}). 

Next we consider warm dense Carbon Hydrogen mix (CH, 1:1). Fletcher \textit{et. al.} measured X-ray scattering ($q=8.43\,{\mathring{\text{A}}}$) of shock compressed CH.\cite{Fletcher2014} It is estimated that the CH reached $6.05$ $\text{g/cm}^3$ (assuming an average ionization of $4$ for C and $1$ for H) and $T= 10$ eV. Here our KS calculations utilize a frozen 1s core pseudo potential, while our OF calculations use a local regularization potential with all electrons treated as valence. We again see the produces a reasonable plasmon peak position, but its plasmon oscillation does not decay. Qualitatively the TFW+D produces a reasonable, but underdamped, plasmon response compared to Kohn Sham. As expected, we see that the frozen core KS result decays rapidly at high frequencies. This is precisely the regime where the carbon 1s excitation are important, $\gtrapprox 400$ eV. The inclusion of all electrons in the TD-OF-DFT does then produce a better response, compared to previous calculation and experiment, in this high frequency regime.

\begin{figure}[t]
    \centering
    \includegraphics[width=0.85\linewidth]{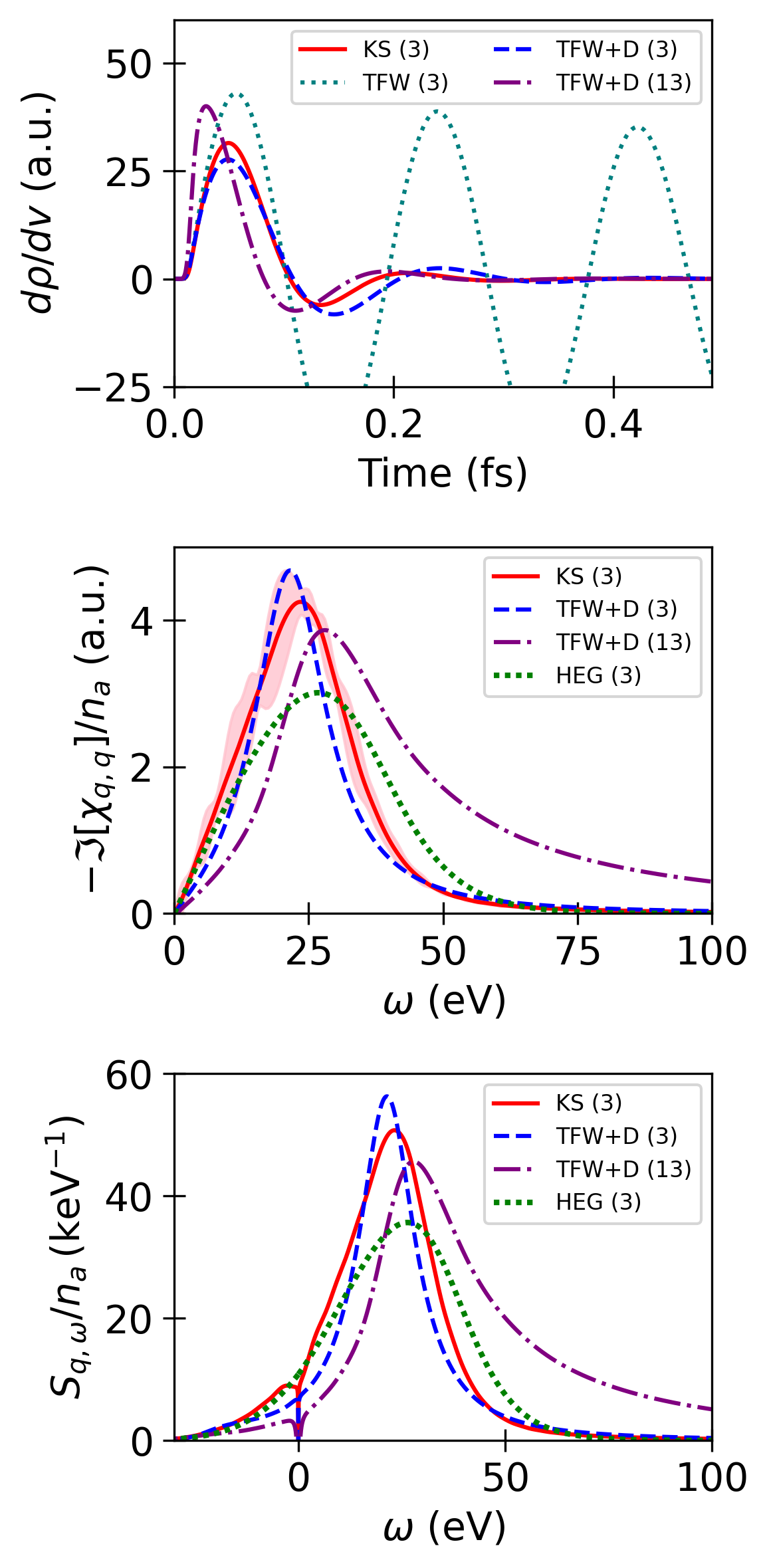}
  \caption{ Solid density Aluminum at $\rho=2.7$ $\text{g/cm}^3$ at $T=6$ eV. Same format as Fig.\ref{fig:Be_1}. Results are averaged over five molecular dynamics snapshots with 32 atoms.  Results are calculated using Kohn Sham (KS (3), red solid), Orbital Free at the Thomas Fermi von-Weizsäcker level (TFW (3), teal dotted), and Orbital Free with TFW plus dynamic potential (TFW+D (3), blue dashed), each using a 3 valence electron pseudopotential and a 13 valence electron TFW plus dynamic potential (TFW+D (13), purple dash-dot). For the frequency domain we do not plot the TFW result, but instead the Linhard response and resulting DSF for 3 free-electrons per atom (HEG (3), green dense-dotted).}
    \label{fig:CH}
\end{figure}

Finally, we consider the experimental conditions reached by Sperling \textit{et. al.}, Aluminum at solid density and $T=6$ eV.\cite{Sperling2015} For our Kohn Sham calculations we utilize a 3 valence, 8 frozen core electrons pseudopotential, while in our OF calculations we utilize both a 13 valence electron regularization potential and a ``bulk local pseudopotential'' produced by the Carter group with only 3 valence electrons.\cite{Zhou04} This pseudo-potential is developed for orbital-free DFT in order to accurately capture the effects of the frozen core electrons. Using the 3 valence electron pseudopotentials, we can directly compare the KS and OF (TFW+D) result for ``nearly free" electrons isolated from core orbitals. We see that the frequency up-shift of the TFW+D is removed and the decay rate / spectral width is much improved compared to the previous cases. For the 13 valence electron regularization potential the frequency up-shift is present ant the long tail at higher frequencies is present. Unlike the previous CH case, here the response is occurring at much smaller frequencies, and the plasmon response should be well-separated from any bound-free contributions. This is verified in the experimental result.\cite{Sperling2015} Thus the 3 valence electron results are more physically relevant. We also consider the random phase approximation response of the homogeneous electron gas, at the density corresponding to 3 free-electrons per Al. We see that the KS and TFW+D produce the energy downshift and sharp decay at higher frequencies which was seen experimentally, see Fig. 2a of reference \cite{Sperling2015}.

\section{Discussion}

From these four systems, which cover a wide range of $q$ and $\omega$ space we see some common themes. The inelastic DSF, experimentally measured by XRTS experiments, is much more sensitive to the electronic structure than stopping power. We previously compared TD-DFT calculations of stopping powers at the KS, TFW and TFW+D levels.\cite{White2018,White2022,Malko22} However, stopping power essentially depends on integration of the density response function over frequencies, with the upper limit of that integration depending on the projectile velocity.\cite{White2022} Thus at high projectile velocities the stopping power becomes largely insensitive to the details of the density response. TFW+D was shown to improve results below and near the ``Bragg Peak" but had little effect at high velocities, though the numerical stability of the time-dependent propagation is improved.\cite{White2018} Since XRTS provides frequency dependent data, it more directly probes the density response. We see that the DSF is much more sensitive to theory which calculates the electron response. Thus in the development of both Kohn Sham and Orbital free functionals for warm dense matter, XRTS is and will continue to be a critical experimental benchmark.   

There are two areas where we can hope to improve the TD-OF-DFT results, the treatment (or lack-there of) the bound electrons and improved dynamic functionals to better capture the plasmon response. We see from the Aluminum case, that if the plasmon response is isolated from the bound-free response, and the core electrons can be frozen, then the TD-OF-DFT and TD-KS-DFT results agree fairly well. Potentially, non-local or higher order kinetic energy functionals could improve on the remaining discrepancies. But frozen core pseudo-potentials are not common for orbital-free DFT. The bulk local pseudopotential approach developed by Carter in a promising route, but there are currently only a handful of pseudo-potentials available.\cite{Zhou04} The treatment of the bound-free response remains a challenge for orbital-free, however average-atom / Real space Green's function approaches may be more reliable for bound-free transitions since the bound electron orbitals are less sensitive to inter-atomic effects.\cite{Johnson2012,Mattern2013, Hou2015}  

Finally, we must determine when, if ever, TD-OF-DFT is needed when mixed-stochastic-deterministic DFT is an available option. Mixed TD-DFT largely overcomes the computational scaling limitations of system size and temperatures that TD-OF-DFT seeks to address. If resources are available to perform mixed TD-DFT then clearly it should be the preferred choice. It is about one to two orders of magnitude more computationally expensive in terms of compute-core hours.  However, it scales very well on high performance computing machines. Thus, if wall-time is an issue, but compute-core time is not then mixed TD-DFT is a better option. If, however, overall resources are limited, then TD-OF-DFT is a very efficient choice. Additionally, the comparison of OF and KS simulations can be potentially instructive in understanding what are the important physics affecting the response. While mixed TD-DFT comparison to experiment may be done at a single $q$ point, assuming a known density and temperature, TD-OF-DFT is efficient enough to rapidly scan many q-points, densities and/or temperatures. Thus we believe that TD-OF-DFT still has an important role to play in studying warm dense matter response.

\section{Conclusion}
We explored the application of TD-OF-DFT in calculating the inelastic portion of the dynamical structure factor for warm dense matter, comparing it to the more computationally intensive mixed-stochastic-deterministic KS TD-DFT. We tested these approaches across several experimentally relevant WDM systems, including solid-density aluminum, beryllium, and a carbon-hydrogen mixture, and verified our approach against existing compressed beryllium calculations. Our KS results are in good agreement with the experimental results, especially when considering uncertainty in the temperature and instrument response.
We showed that the dynamic kinetic energy potential is critical for TD-OF-DFT to capture even qualitative plasmon dynamics. However, TD-OF-DFT struggles with bound electron contributions and mixes bound electron response into the plasmon response. This can be overcome by utilizing frozen core pseudopotentials, potentially in future combination with average atom bound-free contributions. However frozen core pseudopotentials for orbital free applications are limited.  Mixed TD-DFT, though more resource-intensive, provides greater accuracy, especially in distinguishing bound and free electron effects. The results highlight the trade-offs between computational efficiency and precision in WDM modeling, suggesting that TD-OF-DFT can still play a valuable role for rapid parameter scanning when resources are limited, while KS TD-DFT remains the preferred choice for high-fidelity simulations.

\begin{acknowledgments}
This work was supported by the U.S. Department of Energy through the Los Alamos National Laboratory (LANL). Research presented in this article was supported by the Laboratory Directed Research and Development program, projects number 20230322ER and 20230323ER. This research used computing resources provided by the LANL Institutional Computing program. Los Alamos National Laboratory is operated by Triad National Security, LLC, for the National Nuclear Security Administration of U.S. Department of Energy (Contract No. 89233218CNA000001).
\end{acknowledgments}

\bibliography{main.bib}

\end{document}